\documentclass[11pt,a4paper]{article}
\usepackage[utf8]{inputenc}
\usepackage{a4wide}
\usepackage{amsmath}
\usepackage{graphicx}
\usepackage{authblk}
\usepackage{color}

\begin{document}
\title{Polymer escape from a confining potential}
\date{\today}

\author[1,2]{Harri M\"okk\"onen}
\author[1,3]{Timo Ikonen}
\author[1,2,4]{Hannes J\'onsson}
\author[1,4]{Tapio Ala-Nissila}

\affil[1]{\footnotesize{Department of Applied Physics and COMP CoE, Aalto University School of Science, P.O. Box 1100, FIN-00076 Aalto, Espoo, Finland \footnote{harri.mokkonen$@$aalto.fi}}}
\affil[2]{Faculty of Physical Sciences, University of Iceland, Reykjav{\'\i}k, Iceland}
\affil[3]{VTT Technical Research Centre of Finland, P.O. Box 1000, FI-02044 VTT, Finland}
\affil[4]{Department of Physics, Brown University, Providence RI 02912-1843, U.S.A.}

\maketitle

\begin{abstract}
The rate of escape of polymers from a two-dimensionally confining potential well has been evaluated using self-avoiding as well as ideal chain representations of varying length, up to 80 beads. Long timescale Langevin trajectories were calculated using the path integral hyperdynamics method to evaluate the escape rate. A minimum is found in the rate for self-avoiding polymers of intermediate length while the escape rate decreases monotonically with polymer length for ideal polymers. The increase in the rate for long, self-avoiding polymers is ascribed to crowding in the potential well which reduces the free energy escape barrier. An effective potential curve obtained using the centroid as an independent variable was evaluated by thermodynamic averaging and Kramers rate theory then applied to estimate the escape rate. While the qualitative features are well reproduced by this approach, it significantly overestimates the rate, especially for the longer polymers. The reason for this is illustrated by constructing a two-dimensional effective energy surface using the radius of gyration as well as the centroid as controlled variables. This shows that the description of a transition state dividing surface using only the centroid fails to confine the system to the region corresponding to the free energy barrier and this problem becomes more pronounced the longer the polymer is. A proper definition of a transition state for polymer escape needs to take into account the shape as well as the location of the polymer.        
\end{abstract}


\section{Introduction}

Polymer translocation is a common process in various biological systems \cite{muthukumar2011polymer}. A better understanding of these processes is important for novel medical applications and treatments as well as for new DNA sequencing technology where the molecule is driven through an artificial channel in a membrane and each nucleotide induces a characteristic currents across the membrane from which it can be identified \cite{Branton2008,Meller2000,Akeson1999,Howorka2001}. Experimental measurements have also shown that it is possible to separate polymers of different lengths and produce an accurate drug delivery system based on translocation \cite{Bonthuis2008,Han1999}. The crossing rate of the polymer can, however, depend strongly on the environment, thereby affecting the measured signal. A better understanding of the translocation dynamics could help make these types of methods more reliable.

A polymer escaping from a metastable external potential well through a narrow channel represents a generic model of such systems. The channel represents both an energetic as well as an entropic barrier. The energy barrier can have contributions from steric effects as well as the electromagnetic field of the channel and ions in the surrounding liquid. The translocation rate of ideal polymers in simple external potentials has been estimated analytically in limits where the chain is either significantly larger or smaller than the external potential well \cite{Park1999,Sebastian2000}. An analogy with semiclassical treatment of quantum tunneling of a particle has been used \cite{Sebastian2006}. Other types of polymers, such as ring polymers have also been studied \cite{Sung1996,Lee2001a,Lee2001,Debnath2010}.  

For more complex polymers, numerical simulations can provide an estimate of the escape rate. But, since polymer escape is typically a rare event on the time scale of atomic vibrations, a direct numerical solution to the equations of motion ('molecular dynamics', MD) becomes impractical. The timescale difference can amount to many orders of magnitude. The path integral hyperdynamics (PIHD) method \cite{Voter1997,Chen2007} makes it possible to accelerate the escape by applying an artificial bias force, thereby reducing the time interval that needs to be simulated, and then subsequently correcting the calculated rate to give an estimate of the true rate. This methodology has previously been applied to polymer escape from a one-dimensional external potential \cite{Shin2010}. PIHD has been shown to work even for a time dependent bias force \cite{Ikonen2011}. We note that this method is different from Voter's hyperdynamics method \cite{Voter1997}, where the bias potential is designed to vanish at first order saddle points.

Alternatively, an estimate based on statistical mechanics rather than dynamical trajectories started at the initial state can be used to estimate transition rates, if the initial state is assumed to reach and maintain equilibrium distribution of energy in all degrees of freedom. Such a rate theory approach involves much less computational effort than simulation of trajectories. Kramers theory of chemical reactions in solutions \cite{Kramers1940} has, for example, been applied to estimate the escape rate of polymers. It assumes, however, a one-dimensional reaction coordinate and the question is how to define such a coordinate in a system with multiple, coupled degrees of freedom as in a polymer. 

In this article, we present studies of the escape of polymers from a two-dimensional external potential well using standard Langevin dynamics, PIHD, and Kramers rate theory. Two different polymer models are studied: (1) Self-avoiding polymers without bending stiffness, and (2) ideal polymers without excluded volume. The polymers are modeled using the finite extension nonlinear elastic (FENE) model with Lennard-Jones repulsive interaction.  

The article is organized as follows: In the following section, the methodology is presented, including a description of the polymer models, the PIHD method,  Langevin dynamics, and the rate theory.  In section 3, the results are presented, followed by a discussion in Section 4.


\section{Methodology}

\subsection{Polymer models}

\begin{figure}
\centering
\begin{tabular}{cc}
 \includegraphics[scale=0.45]{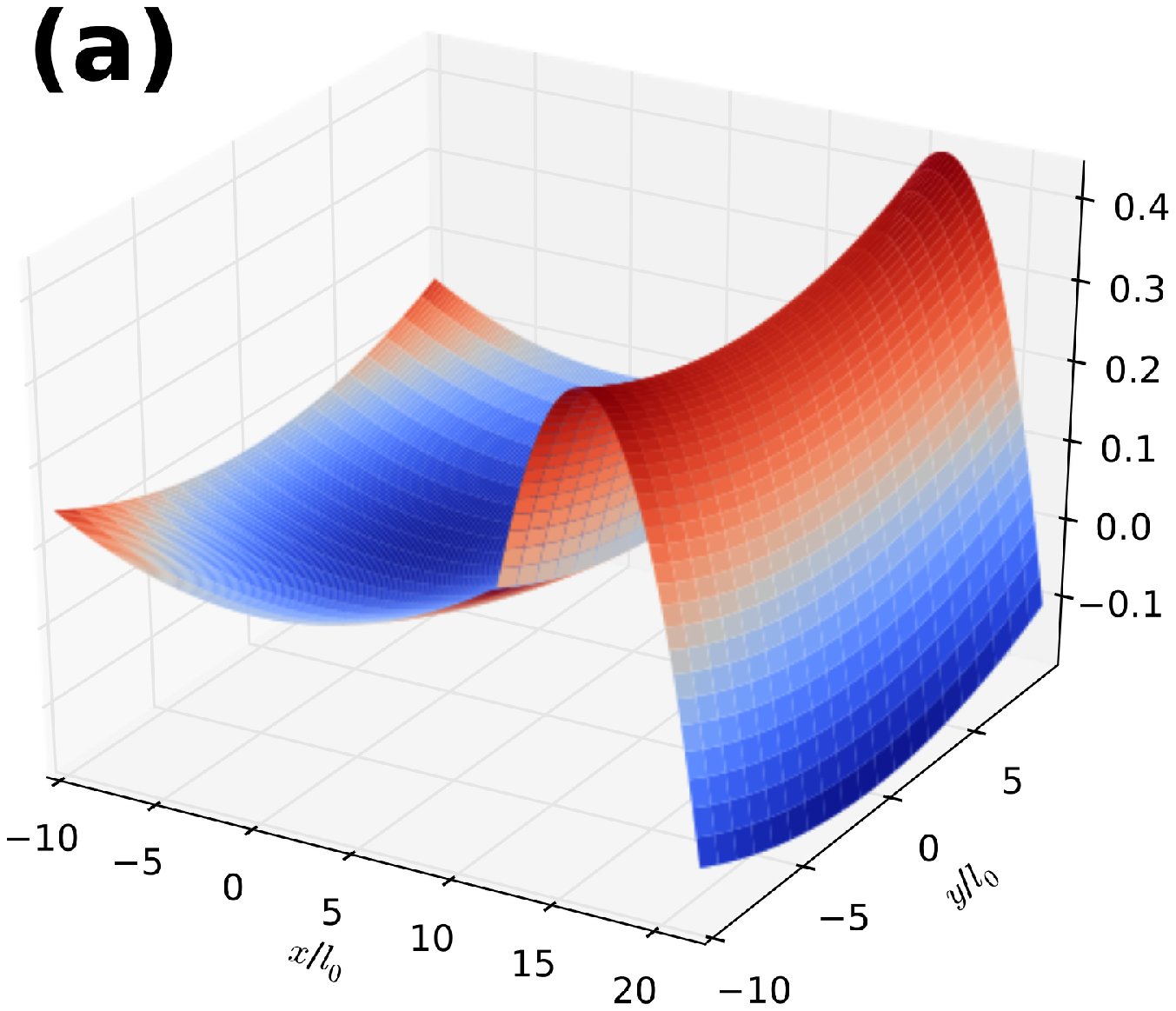} &  \includegraphics[scale=0.34]{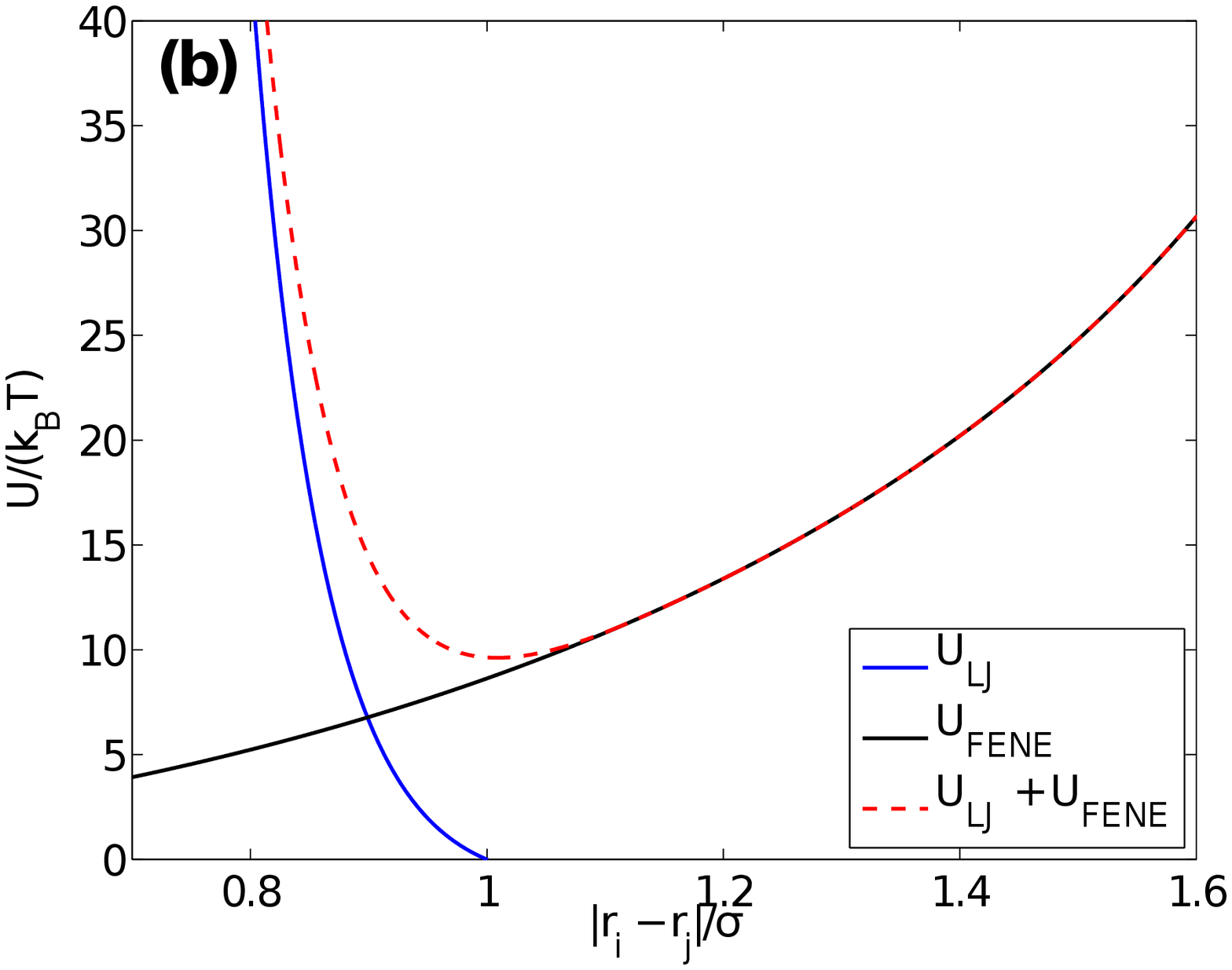} \\
\end{tabular}
\caption{\emph{(a) Illustration of the external potential $V_{\mathrm{ext}}(x,y)$, see Eq. \eqref{eq:ext_potential}. (b) The interaction potential $U_{\mathrm{int}}$ between adjacent beads, see Eq. \eqref{eq:int_potential}. The LJ potential acts between all the beads in the self-avoiding chain model but only between consecutive beads in the ideal chain model. The $U_\mathrm{FENE}$ component of the potential diverges at $R_0 = 2.0$, setting a maximum in the separation between consecutive beads.} }
\label{fig:potentials}
\end{figure}

The polymers were modeled as strings of beads coupled with an interaction potential and subject to a two-dimensional external potential. The Hamiltonian is 
\begin{equation}
\mathcal{H}(\{\mathbf{r}_i, \mathbf{v}_i\}) = \sum_{i=1}^N \frac{m}{2}|\mathbf{v}_{i}|^2 + \Phi(\{\mathbf{r}_i\}), 
\label{eq:tsthamiltonian}
\end{equation}
where $m$ is the mass of a bead and $N$ is the number of beads in the polymer. The position of bead $i$ is given by $\mathbf{r}_i$ and the velocity by $\mathbf{v}_{i}$. The interaction potential is
\begin{equation}
\Phi(\{\mathbf{r}_i\} ) = \sum_{i=1}^N V_{\mathrm{ext}}(\mathbf{r}_i)  + U_{\mathrm{int}}(\{\mathbf{r}_i\}), \label{eq:potentialpart}
\end{equation}
where the external potential is
\begin{eqnarray}
  V_{\mathrm{ext}}(x,y) &=& \left\{
  \begin{array}{l l}
    \frac{1}{2} \omega_0^2 (x^2 + y^2), & \quad x \leq x_0 ; \\
    \Delta V - \frac{1}{2} \omega_b^2 (x - x_b)^2  + \frac{1}{2}\omega_0^2 y^2,  & \quad  x > x_0. \label{eq:ext_potential}
  \end{array} \right. 
\end{eqnarray}
The quantities $\omega_0$ and $\omega_b$ give the curvature of the well and of the barrier, respectively, $\Delta V$ the height of the barrier located at  $x_b$ and the $x_0$ the cross-over point between the two parabolas. The external potential is illustrated in Fig. \ref{fig:potentials}(a). A potential function with the same $x$-dependence but without confinement in the $y$-direction was used in the work of Shin \emph{et al.} \cite{Shin2010}. 

The interaction between the beads is given by 
\begin{equation}
U_{\mathrm{int}}(\{\mathbf{r}_i\}) = \sum_{i}^{N-1} U_\mathrm{FENE}(|\mathbf{r}_{i}-\mathbf{r}_{i+1}|) + 
\sum_{\langle i,j \rangle}^{N} U_\mathrm{LJ}(|\mathbf{r}_{i}-\mathbf{r}_{j}|),
\label{eq:int_potential}
\end{equation}
where
\begin{equation}
U_\mathrm{FENE}(r) = - \frac{1}{2} k_F R_0^2 ~ \ln (1-r^2/R_0^2), 
\label{eq:fenepotential} 
\end{equation}
and
\begin{equation}
U_\mathrm{LJ}(r) =   4 \epsilon [(\sigma/r)^{12} - (\sigma/r)^6 + 1/4]. 
\label{eq:ljpotential}
\end{equation} 
The repulsive interaction between the beads is a Lennard-Jones (LJ) potential that is truncated and shifted so that $U_{LJ}(r) = 0$ if $r > 2^{1/6} \sigma$. The shift by $\epsilon$ ensures continuity of the function. The interaction potential between adjacent beads, which is illustrated in Fig. \ref{fig:potentials}(b), also includes an attractive interaction, the so-called finite extension nonlinear elastic (FENE) interaction. Non-adjacent beads only repel each other through the $U_{LJ}$ potential, the second sum in Eq.  \eqref{eq:int_potential}  then including all pairs of beads. We will refer to this full interaction model as the 'self-avoiding' polymer. For comparison, we have also carried out simulations with a simpler model where non-adjacent beads do not interact at all. The summation over $U_{LJ}$ in Eq.  \eqref{eq:int_potential} then includes only adjacent beads. We will refer to this simpler model as the 'ideal polymer'.

\subsection{Dynamics}

The dynamics of the polymer is given by the Langevin equation
\begin{equation}
m \ddot{\mathbf{r}}_i(t) + \gamma \dot{\mathbf{r}}_i(t) + \nabla_i \Phi (\{\mathbf{r}_i\}) = \mathbf{\Xi}_i (t), \label{eq:langevinmanypart}
\end{equation}
where $\gamma$ is the friction coefficient, $\Phi(\{\mathbf{r}_i\})$ is the total potential energy given by  Eq. \eqref{eq:potentialpart}, $\nabla_i$ is the gradient taken w.r.t the coordinates of the $i$th bead and $\dot{\mathbf{r}}_i$ is the velocity of bead $i$. A Gaussian random force $ \mathbf{ \Xi}(t)$ describes the effect of collisions by solvent molecules and is defined in such a way that $\langle \mathbf{\Xi}(t) \rangle = 0$ and $\langle \Xi_\mu(t) \Xi_\nu(t') \rangle = 2 \gamma k_B T \delta_{\mu,\nu} \delta(t-t')$. Here $ \langle \dots \rangle$ denotes the ensemble average, $\mu$ and $\nu$ Cartesian coordinate indices, $k_B$ the Boltzmann constant, $T$ the temperature, $\delta(t)$ the Dirac delta function and $\delta_{\mu,\nu}$ Kronecker's delta.

The escape rate is defined as the derivative of the escape probability $\mathcal{R} = dP_{\mathrm{esc}} (t)/dt$. For a single particle, the escape probability 
can be written using the path integral formulation as \cite{Chen2007,ikonenthesis} 
\begin{equation}
P_{\mathrm{esc}} (t) = \int_{x_f \geq x_b}  dr \int_{x_0 \leq x_b} dr_0 P(r_0)P(r_0, t_0 | r, t), \label{eq:transitionprobability}
\end{equation} 
where $x_b$ is the position of the barrier top, $P(r_0)$ is the Boltzmann distribution of the initial configurations and 
\begin{equation}
P(r_0, t_0 | r, t) = C \int [Dr] \exp(-\beta I[r(t)]), \label{eq:propabilitydensity}
\end{equation}
is the probability that the particle has moved from $r_0$ at time $t_0$ to $r$ at time $t$. $\int [Dr]$ refers to integration over all possible paths between 
$r_0$ and $r$, $C$ being the normalization constant and $I[r(t)]$ being the action of the path
\begin{equation}
I[r(t)] = \int dt [m \ddot{r}(t) + \gamma \dot{r}(t) + \nabla_i \Phi (r(t))]^2
\end{equation}

For a polymer, an escape event is defined to have occurred when the $x$-coordinate of the centroid (C) of the polymer, $x_\mathrm{C} = (1/N) \sum_{i=1}^N x_i$, has advanced well beyond the location of the barrier maximum, beyond $x=x_b+4$. Eq. \eqref{eq:transitionprobability} can be evaluated numerically as 
\begin{equation}
P_{\mathrm{esc}} (t) = \frac{1}{N_\mathrm{traj}} \sum_{i=1}^{N_\mathrm{traj}} \Theta(t - t_i).\label{eq:transtionprobabilitydistributionnobias}
\end{equation}
where $\Theta(\dots)$ is the Heaviside function, $t_i$ is the escape time of the $i$th trajectory and $N_\mathrm{traj}$ is the total number of simulated trajectories. 

In PIHD an artificial bias potential $V_b(x)$ is added to the external potential. The action can then be split into two parts $I[r(t)] = I_b[r(t)] + I_\Xi[r(t)]$, where $I_b[r(t)]$ is the action for the system in the presence of the bias potential and   
\begin{equation}
I_{\Xi[r(t)]}  = \frac{1}{4 \gamma} \int_{t_0}^t dt' \nabla V_{\text{b}}(\mathbf{r}) \cdot [\nabla V_{\text{b}}(\mathbf{r}) -2\mathbf{\Xi}(t')], \label{eq:correctionfactormanypart}
\end{equation}
is the PIHD correction factor for each trajectory. For the biased system, Eq. \eqref{eq:transtionprobabilitydistributionnobias} becomes 
\begin{equation}
P_{\mathrm{esc}} (t) = \frac{1}{\mathcal{N}_\Xi} \sum_{i=1}^{N_\mathrm{traj}} \Theta(t - t_i) \exp(-\beta I_\Xi[r_i(t)]), \label{eq:transtionprobabilitydistribution}
\end{equation}
where $t_i$ is the escape time of trajectory $i$ \cite{Shin2010} and $N_\mathrm{traj}$ is the total number of simulated trajectories [20, page 25]. Eq. \eqref{eq:transtionprobabilitydistribution} gives the transition probability for an unbiased system in terms of crossing probability obtained from trajectories of the biased system. The normalization factor is
\begin{equation}
\mathcal{N}_\Xi  =   \sum_{i=1}^{N_\mathrm{traj}} \exp({-\beta I_\Xi[r_i(t)]}).\label{eq:pihdnormalisation}
\end{equation}
The bias potential for each bead was chosen here to be $V_b(x) = \frac{1}{2}b \omega_0^2 x^2$ when $x \leq x_0$ and $V_b(x) = -b\Delta V + \frac{1}{2}b \omega_b^2 (x - x_b)^2$ when $ x > x_0$, where $b$ is a parameter to be chosen between $0 \leq b < 1$. Thus the bias potential flattens the external potential along $x$-axis making the escape events more frequent. We tried a few different choices of the bias potential, including a constant force on all the beads as well as dragging the chain from one end. The one chosen here worked best.

An equilibrium distribution for the initial state, $P(\mathbf{r}_0)$, was generated by letting the system thermally relax without bias.  Configurations were then drawn from this equilibrium distribution and the bias potential turned on to generate escape trajectories. Configurations were sampled at time intervals of $2\tau$ where $\tau$ is the relaxation time  \cite{Kopf1997}.


\subsection{Rate theory}

Kramers theory is frequently used to obtain estimates of transition rates for molecules in solution  \cite{Kramers1940,Hanggi1990}. It is based on a Langevin description of the dynamics and different expressions for the rate are obtained depending on the magnitude of the friction coefficient. In the high friction limit,
the Kramers estimate of the crossing rate of a particle escaping from a metastable potential is 
\begin{equation}
\mathcal{R}_{\mathrm{K}} = \frac{\omega_0\omega_b}{2 \pi \gamma_K} e^{-\beta  \Delta V},
\label{eq:kramers}
\end{equation}
where $ \Delta V$ is the height of the energy barrier, $\omega_0$ is the curvature of the energy surface at the initial state minimum and
$\omega_b$ is the magnitude of the negative curvature at the barrier top.

In order to apply the Kramers formula in the present case, the multiple degrees of freedom of the polymer need to be reduced to a single reaction coordinate.  One possibility is to choose the $x$-coordinate of the centroid as the independent variable. An effective potential energy curve for this one degree of freedom is then obtained by thermally averaging over all the other degrees of freedom. The thermal average of a function $f(\mathbf{r})$ for a fixed value of the $x$-coordinate of the centroid is
\begin{equation}
\langle f \rangle_\mathrm{C} =  \frac{1}{Z_N(x_\mathrm{C})}\int \prod_{i=1}^N
 d\mathbf{r}_i' f(\{\mathbf{r}_i'\}) \delta(x_\mathrm{C} - \underbrace{\frac{1}{N}\sum_{j=1}^N x'_j}_{=x_\mathrm{C}^\prime}) 
 e^{-\beta \Phi(\{\mathbf{r}_i'\})},
 \label{eq:effpotential}
\end{equation}
where 
\begin{equation}
Z_N(x_\mathrm{C}) = \int \prod_{i=1}^N
 d\mathbf{r}_i'  \delta(x_\mathrm{C} - \frac{1}{N}\sum_{j=1}^N x'_j) 
 e^{-\beta \Phi(\{\mathbf{r}_i'\})}.
\end{equation}

By applying this averaging to the total potential, an effective energy curve $\Phi_{\mathrm{eff}}(x_\mathrm{C},N) = \langle \Phi \rangle_\mathrm{C}$ is obtained. The friction coefficient for this reduced dimensionality system is $\gamma_\mathrm{eff} \approx N \gamma_K$. The effective friction coefficient in the Kramers rate expression, $\gamma_K$, was adjusted here to obtain a good estimate of the simulated escape rate of a single bead and turned out to be $\gamma_K = 0.82 \gamma$. A Kramers approximation for the escape rate of a polymer with $N$ beads is thus obtained as
\begin{equation}
\mathcal{R}_{\mathrm{K}}(N) = \frac{\omega_{0,{\mathrm{eff}}}(N) \omega_{B,\mathrm{eff}}(N)}{2 \pi \gamma_\mathrm{eff}} e^{-\beta \Delta E_{\mathrm{eff}}(N)}. 
\label{eq:crossratekramers}
\end{equation}
From the shape of the effective potential curves, the parameters $\omega_{0,{\mathrm{eff}}}(N)$, $\omega_{B,\mathrm{eff}}(N)$ and $\Delta E_{\mathrm{eff}}(N)$ were estimated by fitting parabolas at the initial state minimum and at the barrier maximum.  

The internal degrees of freedom of the polymer contribute to the effective potential curve $\Phi_\mathrm{eff}$.  Alternatively, an effective external potential curve without including the interaction between beads can be calculated as $V_{\mathrm{eff}}(x_\mathrm{C},N)  = \langle V_\mathrm{ext} \rangle_\mathrm{C}$. We compare below the two energy curves and the Kramers rate estimates obtained from each one.


\subsection{Simulation parameters}

The values of the various parameters used in the simulations were $m=m_0=1870$ amu, $k_BT = 1.2$ and $\sigma = 1.02$ nm, which corresponds roughly to three base pairs of DNA. These parameters fix the mass, length and energy scales resulting in a time scale characteristic of the LJ potential as $t_\mathrm{LJ} = \sqrt{m \sigma^2 / \epsilon} = 30.9$ ps, where $\epsilon = 1$ $k_B T$. The external potential was defined by parameters $\omega_0 = 0.0014$, $\omega_b = 0.032$, $\Delta V = 0.3$ $k_B T$. The barrier was located at position $x_b  =16$ and the crossover between the two parabolas at $x_0 = 12$. The parameter in the FENE spring constant was $k_F = 15$ and the maximum FENE separation $R_0 = 2.0$. The Langevin equation was integrated in time using a velocity Verlet type algorithm \cite{Brunger1984} which is particularly well suited for PIHD. The effective potential curves $\Phi_\mathrm{eff}$ and  $V_\mathrm{eff}$ were sampled during the escape simulations. The PIHD bias parameter $b$ was chosen between $0.7\dots0.9$.

The chosen friction in the Langevin dynamics, $\gamma = 0.7$ ($=3.8 \times 10^{-6}$ kg/s), corresponds to the high friction range of Kramers' theory $\gamma \gg \pi \omega_b$ \cite{Kramers1940}. With this choice of friction, viscosity of the fluid surrounding polymer can be estimated to be $\eta \approx 1.3 \times 10^{-5}$ g(cm s)$^{-1}$ (for water $\eta = 1$ g (cm s)$^{-1}$)  \cite{ollila2011}.


\section{Results}

\begin{figure}
\centering
\includegraphics[scale=0.37]{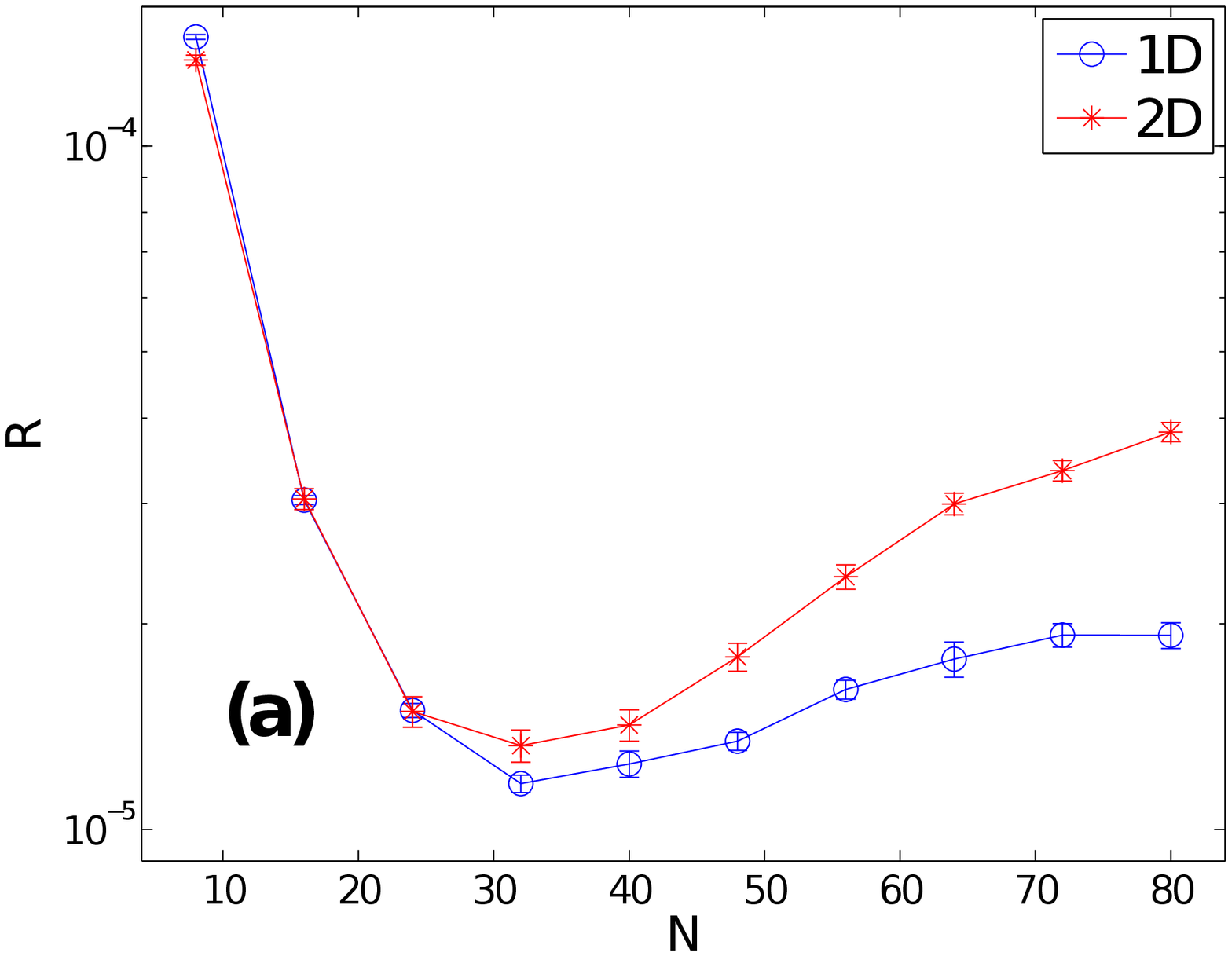} 
\includegraphics[scale=0.37]{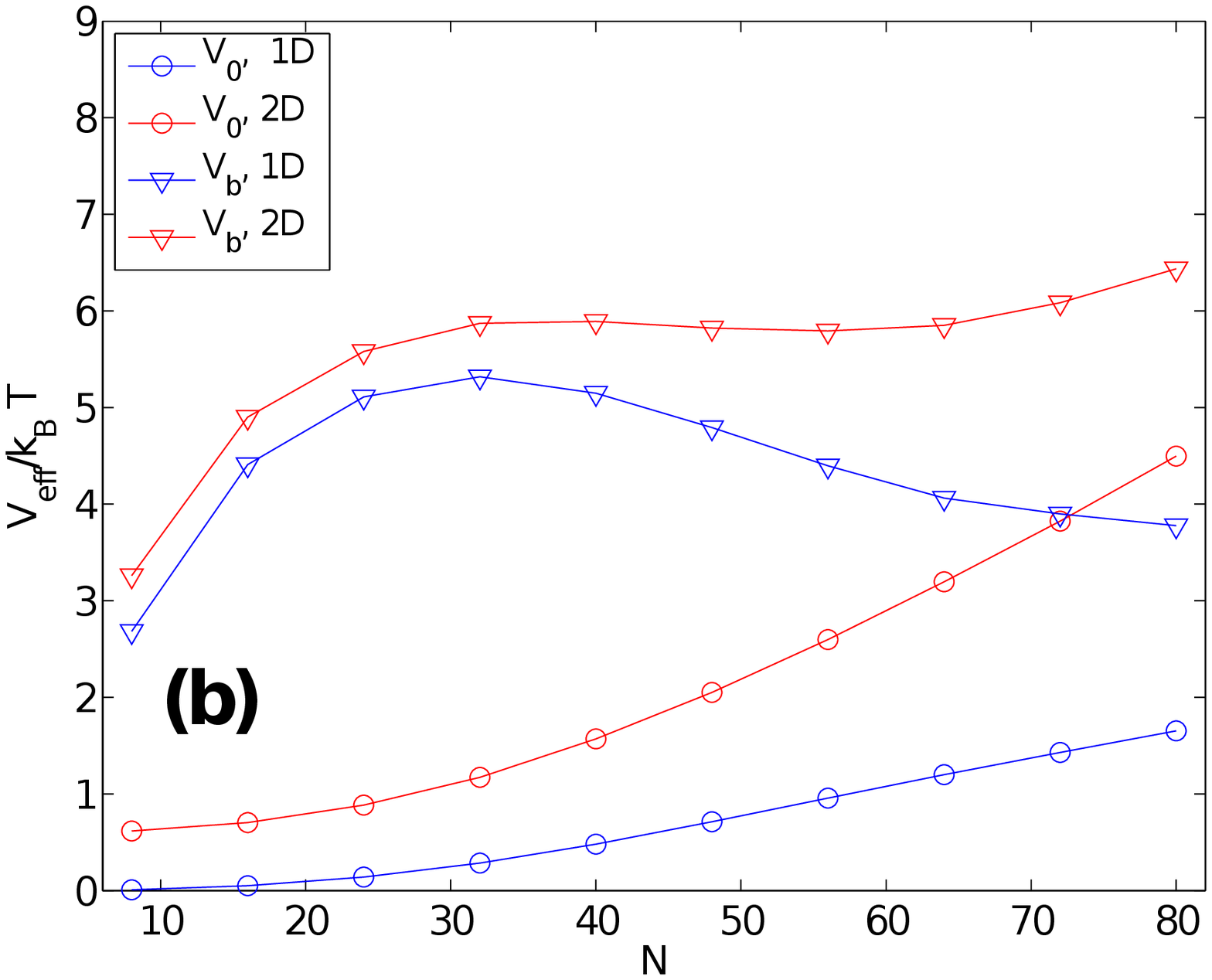} 
\caption{\emph{(a) Comparison of the escape rate of self-avoiding polymers from one- and two-dimensionally confining potentials.  
Red line: A potential well with two-dimensional confinement by Eq. \eqref{eq:ext_potential}. Blue line: A potential well with with the same $x$ dependence but no confinement in the $y$-direction (taken from \cite{Shin2010}). The added confinement in the two-dimensional well results in enhanced escape rates for the longer polymers. (b) Contribution of the external potential to the energy barrier in the potential with confinement in $y$-direction (red lines) and without (blue lines). Circles present the effective external potential $V_\mathrm{eff}(x_\mathrm{C})$ in the well bottom $x_\mathrm{C} = 0$ and triangles at the barrier top $x_\mathrm{C} = x_b$.}}
\label{fig:rates1d2d}
\end{figure}

The escape probability was calculated for polymers with up to $N=80$ beads using PIHD and Eq. \eqref{eq:transtionprobabilitydistribution}. A linear least squares fit to the calculated $P_{\mathrm{esc}} (t)$ was then used to estimate the escape rate $\mathcal{R}(N)$. A comparison of the present simulations using a two-dimensionally confined external potential and previously reported simulations \cite{Shin2010} for an external potential without confinement in $y$-direction (similar as Eq. \eqref{eq:ext_potential} but with no terms depending on $y$) is shown in Fig. \ref{fig:rates1d2d}. The added confinement in the two-dimensional well results in enhanced escape rate for the longer polymers and compared to the one-dimensional case. A pronounced minimum in the rate is observed at around $N=30$. This occurs because the repulsive interactions between beads in the longer polymers raise the free energy of the initial state with respect to the transition state.  

The total energy depends strongly on chain length with lower bound $\Phi_\mathrm{eff} > 9(N-1)$ but the values of $V_\mathrm{eff}$ in Fig. \ref{fig:rates1d2d} reveals that when two-dimensional confinement is present $V_0 = V_\mathrm{eff}^\mathrm{ext}(x_0)$ increases faster. This is due to "crowding" in the well; the longer self-avoiding chains cannot fit into the initial state any more so they experience much higher  external potential. In the potential without confinement, the energy barrier $V_\mathrm{eff}(x_b)$ starts to decrease after $N=32$ corresponding the rate minimum which implicates that the chain is more elongated over the barrier when $y$-directional confinement is not present.

\begin{figure}
\centering
\begin{tabular}{ll}
\includegraphics[scale=0.37]{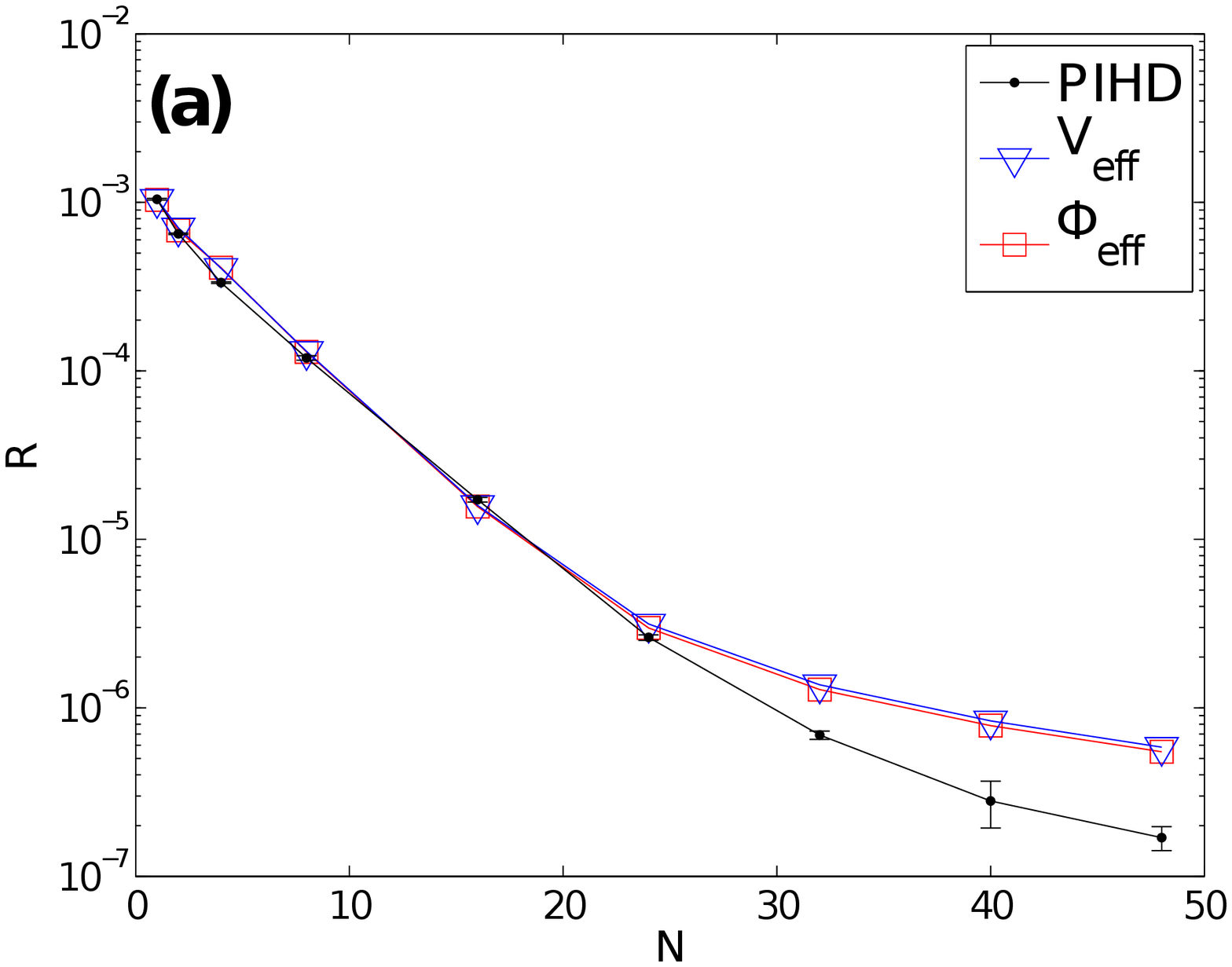} & \includegraphics[scale=0.37]{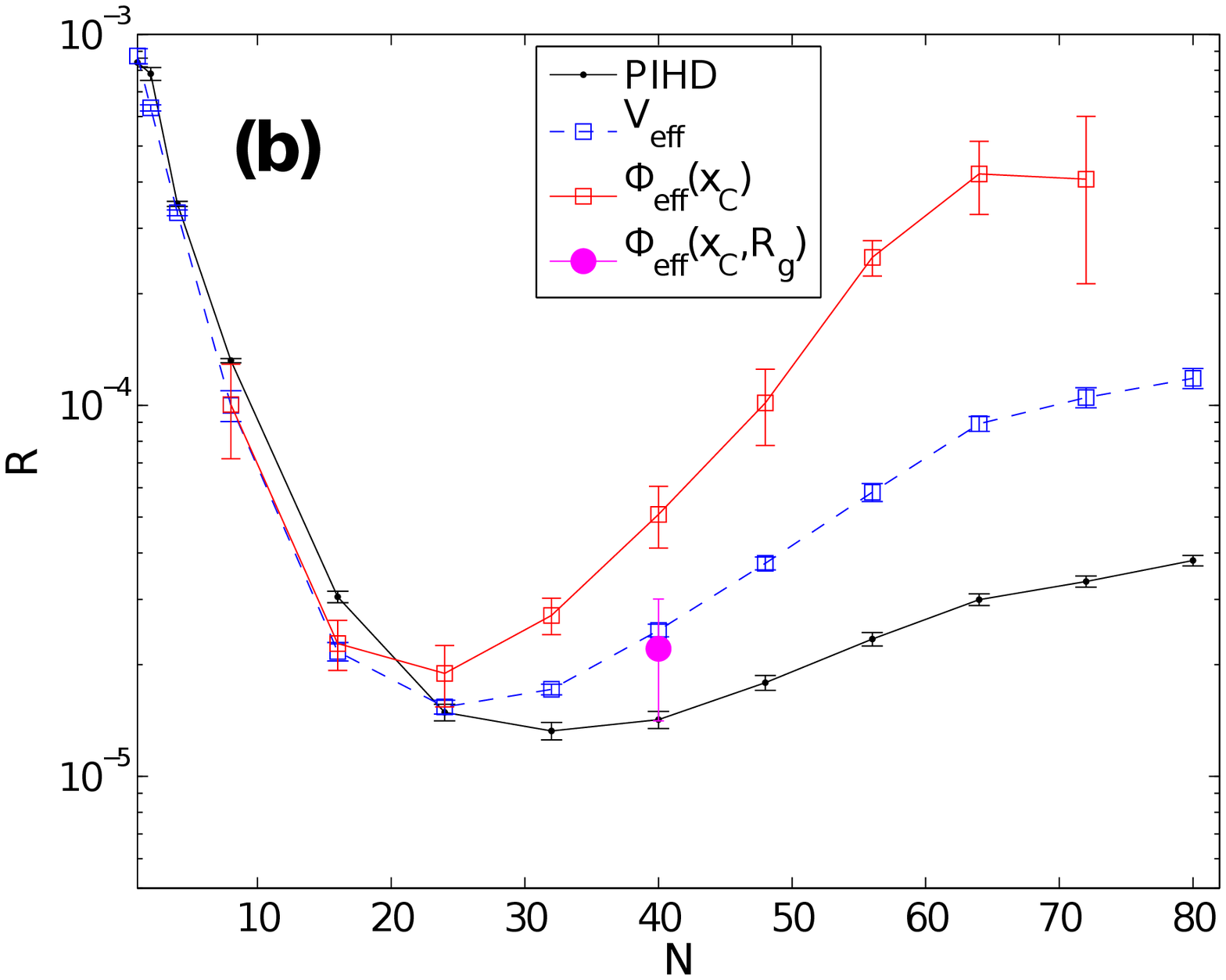} \\
\end{tabular}
\caption{\emph{Escape rate for (a) ideal polymers, and (b) self-avoiding polymers calculated using PIHD and estimated using Kramers theory, Eq. \eqref{eq:crossratekramers}. The escape rate of the ideal polymers is monotonically decreasing as a function of $N$ up to the maximum length simulated, while the escape rate of self-avoiding polymers exhibits a minimum around $N=30$. The purple dot presents the rate computed using the energy barrier averaged over the tilted line in Fig. \ref{fig:effsurf40}. It shows that taking the shape of the polymer into account in reaction coordinate improves the rate given by Kramers theory. }}
\label{fig:rates}
\end{figure}

When the repulsive interaction between non-adjacent beads is turned off (no excluded volume interactions), in the ideal polymers, see Fig. \ref{fig:rates}, this minimum disappears. In this case, the escape rate continues to drop past $N=30$. This shows that the reason for the minimum observed in the escape rate of the self-avoiding polymers is due to the repulsive interactions between non-adjacent beads. The simulations of the ideal polymers were carried out up to $N=48$, beyond which the free energy barrier becomes so large that even with PIHD the simulations become excessively long.

\begin{figure}
\centering
\includegraphics[scale=0.37]{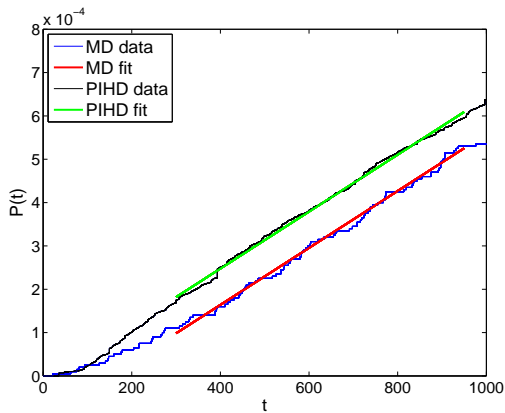} \\
\caption{\emph{Probability that an ideal polymer of length $N=32$ has escaped, $P(t)$, calculated using either  MD or PIHD simulations with $2 \times 10^6$ trajectories. A linear least squares fit to the MD simulation data gives the rate $\mathcal{R} = (6.5 \pm 1.4) \times 10^{-7}$, and a fit to the PIHD simulation the rate $\mathcal{R} = (6.5 \pm 0.9) \times 10^{-7}$.}}
\label{fig:pihd}
\end{figure}

PIHD is, nevertheless, found to be efficient for the ideal chains. MD simulations with $2 \times 10^6$ trajectories and chain length $N=32$ give a root mean square error of $\sigma_\mathrm{MD} = 1.4 \times 10^{-7}$ for the rate, while PIHD with same parameters gives  $\sigma_\mathrm{PIHD} = 0.9 \times 10^{-7}$. Knowing that root mean square error scales as $\sigma \sim N_\mathrm{traj}^{-1/2}$ we can estimate that using MD solely would need approximately 2.5 times more trajectories for same accuracy. For chain length $N=40$ this ratio is approximately 3 while for $N=24$ it is approximately unity illustrating that PIHD is more beneficial for the longer chains. Typical data for $P(t)$ are shown in Fig. \ref{fig:pihd} for straight MD and PIHD simulations for an ideal chain with $N=32$.

\begin{figure}
\centering
\begin{tabular}{ll}
\includegraphics[scale=0.37]{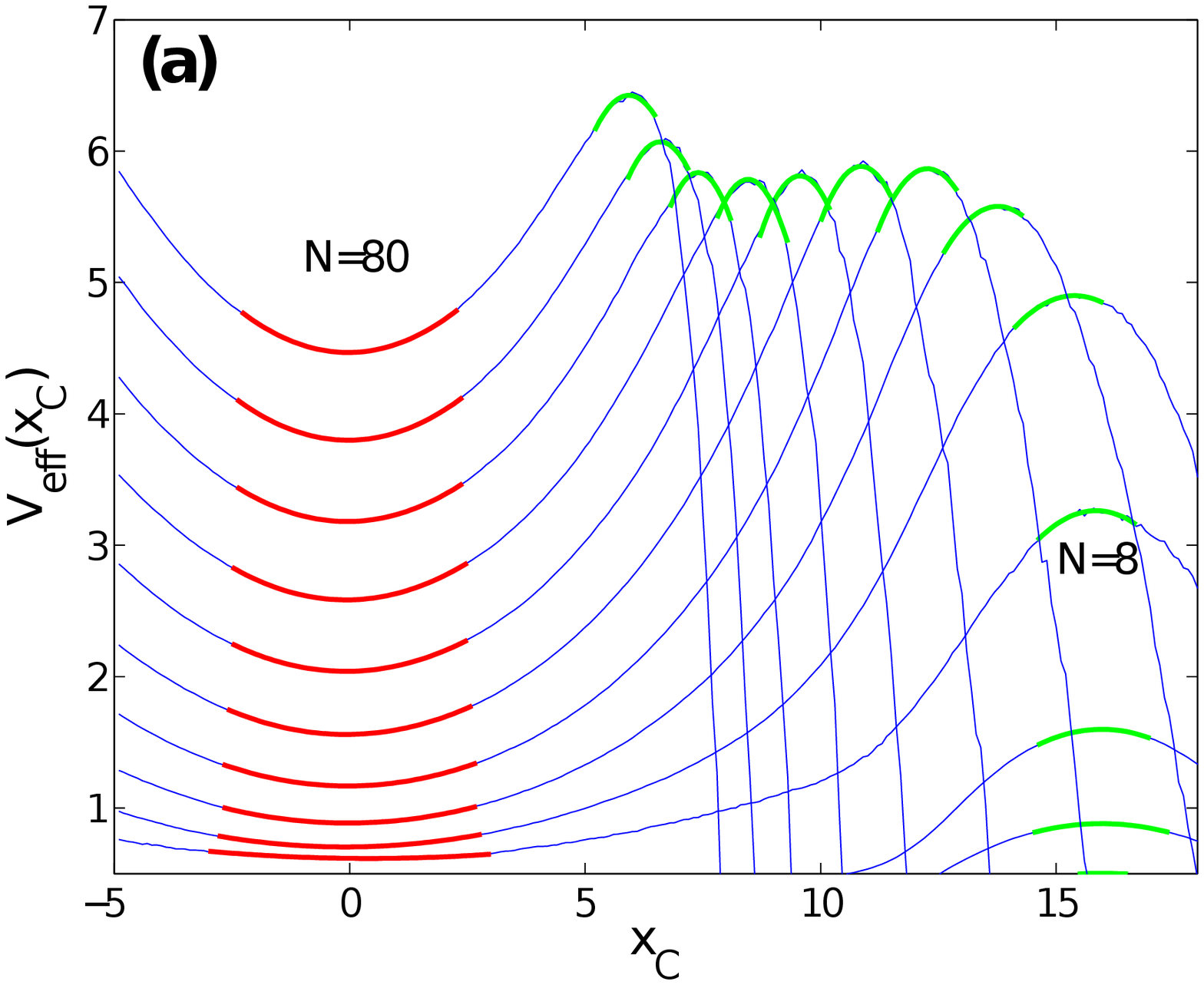} & \includegraphics[scale=0.37]{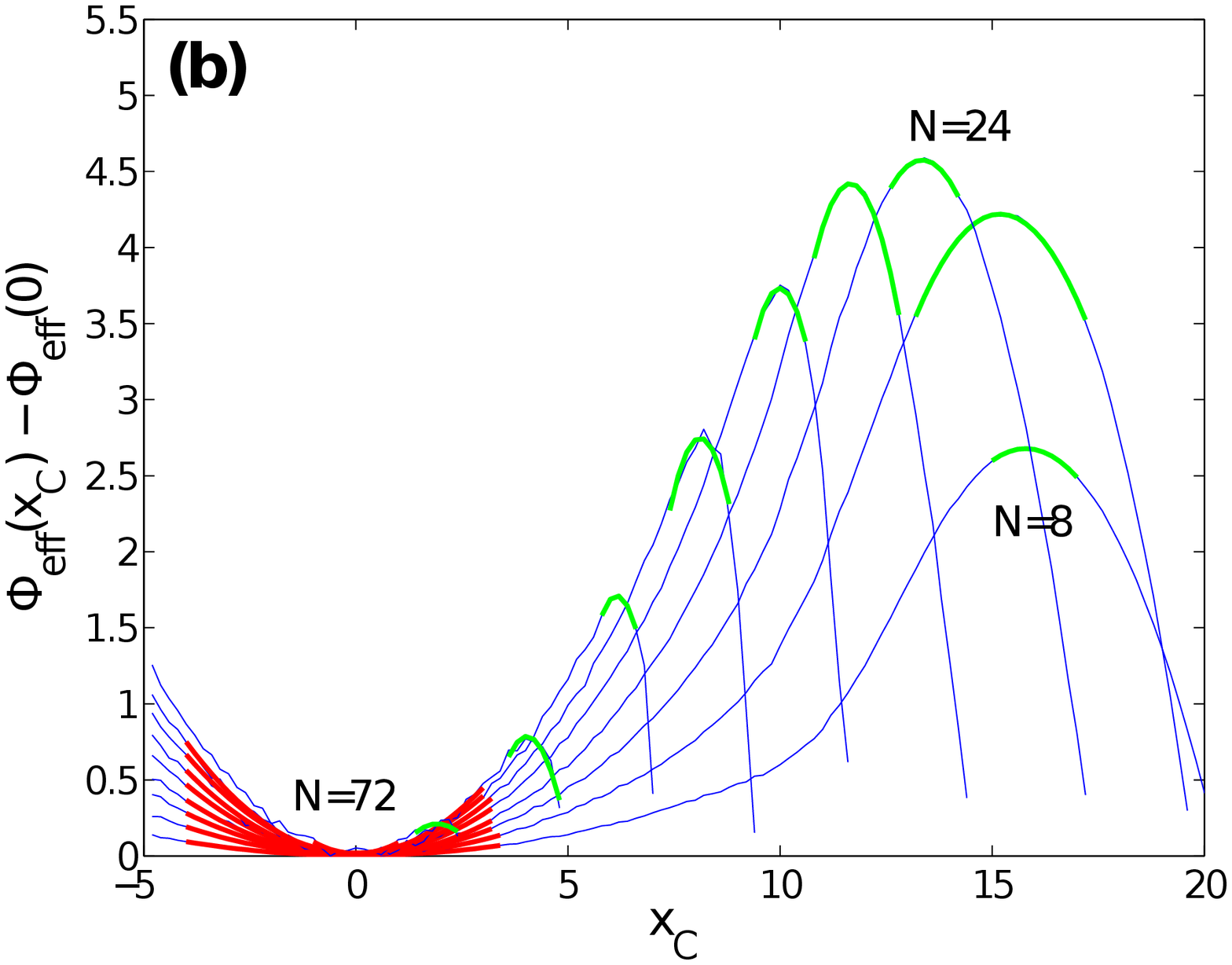} \\
\end{tabular}
\caption{\emph{
Effective potential energy curves for self-avoiding polymer escape from a two-dimensional well, using the $x-$coordinate of the centroid as a reaction coordinate and thermally integrating over all other degrees of freedom, see Eq. \eqref{eq:effpotential}. (a) Effective external potential $V_{\mathrm{eff}}(x_\mathrm{C},N)$. (b) Effective full potential energy $\Phi_{\mathrm{eff}}(x_\mathrm{C},N) - \Phi_{\mathrm{eff}}(0,N)$. Second degree polynomial fits are shown in red and green  for well minima and maxima, respectively. The $\Phi_\mathrm{eff}(x_\mathrm{C},N)$ curves shift upward with $N$ so $\Phi(0,N)$ is subtracted for better illustration.}}\label{fig:sacurves}
\end{figure}

The effective potential curves obtained by fixing the $x$-coordinate of the centroid of the polymer and thermally averaging over the positions of beads according to Eq. \eqref{eq:effpotential} are shown in Fig. \ref{fig:sacurves}. The barrier to the escape of the self-avoiding polymer in the $\Phi_\mathrm{eff}$ effective potential first increases with length and then decreases while the location of the maximum monotonically shifts towards the initial state minimum. This shift is also seen in the barrier of the average external potential $V_\mathrm{eff}$. The statistical sampling is easier for the self-avoiding polymer since the effective energy barrier is lower and the direct dynamics sample the saddle point region better. For $N=80$, the effective barrier for the self-avoiding polymer was too small to obtain good statistics.

An estimate of the escape rate using Kramers rate theory is obtained by fitting the effective potential curves with parabolas to 
extract estimates of the parameters $\omega_{0,{\mathrm{eff}}}(N)$, $\omega_{B,\mathrm{eff}}(N)$ and $\Delta V_{\mathrm{eff}}(N)$  in the rate expression.   

\begin{figure}
\centering
\begin{tabular}{ll}
\includegraphics[scale=0.37]{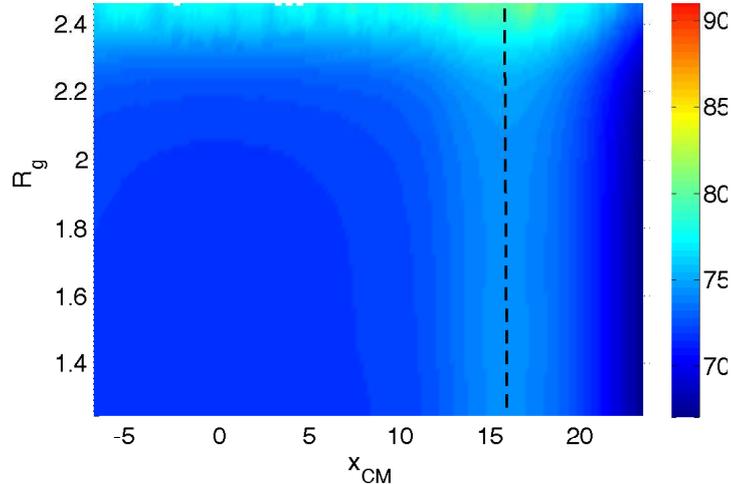} \\
\end{tabular}
\caption{\emph{Contour graph of the effective potential energy surface, $\Phi_{\mathrm{eff}}(x_\mathrm{C},R_g,8)$, for the self-avoiding polymer
where $x_\mathrm{C}$ is the x-coordinate of the centroid and $R_g$ is the radius of gyration of a polymer with 8 beads. In this case the energy barrier lies close to vertical line corresponding to a fixed $x_\mathrm{C}$ (see dashed line). A constraint based on the centroid alone can then be used to define a good reaction coordinate.}}  \label{fig:effsurf8}
\end{figure}

\begin{figure}
\centering
\begin{tabular}{ll}
\includegraphics[scale=0.47]{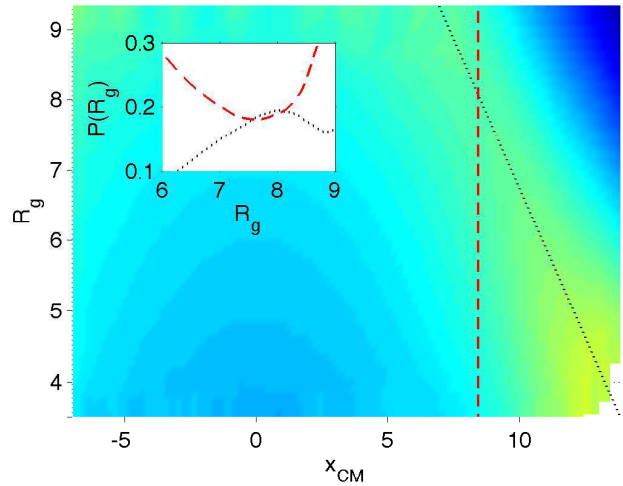} \\
\end{tabular}
\caption{\emph{Contour graph of the effective potential energy surface, $\Phi_{\mathrm{eff}}(x_\mathrm{C},R_g,40)$, for the self-avoiding polymer where $x_\mathrm{C}$ is the x-coordinate of the centroid and $R_g$ is the radius of gyration of a polymer with 40 beads. In this case the energy barrier is titled with respect to a line of a fixed $x_\mathrm{C}$ (dashed line) and a constraint based on the centroid alone does not give a good reaction coordinate. Thermal sampling along the dashed red line is dominated by configurations that either have larger or smaller $R_g$ than the value at the energy barrier (dashed red line in the inset), resulting in an underestimate of the energy barrier. Sampling within the tilted, dashed black line line, however, confines the system within the barrier region and has maximum density at the first order saddle point (dotted black line in the inset) and gives a larger value of the activation energy. This shows that a reaction coordinate using the centroid alone will give an underestimate of the activation energy for polymers of this length and that a good reaction coordinate needs to be defined both in terms of the location and shape of the polymer.}}
\label{fig:effsurf40}
\end{figure}

The rate estimates obtained using Kramers theory Eq. \eqref{eq:crossratekramers} applied to the effective potential curves, are compared with the PIHD simulated results in Fig. \ref{fig:rates}. The rate theory gives  behavior qualitatively similar to the simulations for both polymer models but severely overestimates the escape rate of the longer polymers, especially the self-avoiding ones. The reason for this is an underestimate of the energy barrier to escape in the effective potential curves. The use of the centroid coordinate as a reaction coordinate does not confine the polymers to the transition state region which then leads to an underestimate of the energy barrier. This can be seen by evaluating an effective potential function of two variables, the radius of gyration, $R_g$, as well as the centroid. Such two-dimensional effective potential surfaces are shown in Fig. \ref{fig:effsurf8} for self-avoiding polymer with $N=8$ and in Fig. \ref{fig:effsurf40} for $N=40$.  While the energy ridge for the shorter polymer is aligned with the $x_\mathrm{C}=16$ vertical line, showing that a constraint based on the centroid coordinate alone can confine the system at the barrier, the ridge for the longer polymer is significantly tilted with respect to the vertical axis. This means that a constraint based only on a fixed value of the $x$-coordinate of the centroid cannot constrain the system in the high barrier region. When the thermal averaging of the other degrees of freedom is carried out for the longer polymer and $x_\mathrm{C}=8$, the polymer  either has rather compact configurations with a small value of the radius of gyration, or a significantly larger value. The intermediate values that correspond to the energy ridge are rarely sampled, as shown in the inset of Fig. \ref{fig:effsurf40}. When, however, the effective energy curve is defined by averaging along a line that is tilted in the ($R_g,x_\mathrm{C}$) plane, so as to lie along the energy ridge, the system cannot escape the high energy region and the vicinity of the first order saddle point is sampled, giving a larger average energy barrier to escape.

If the energy is averaged over the tilted line in Fig. \ref{fig:effsurf40} it is $\Delta \Phi' = 4.7$ being higher than $\Delta \Phi = 3.7$ in Fig. \ref{fig:sacurves} where the average is taken over the straight line. Using Eq. \eqref{eq:crossratekramers} we can obtain the corrected estimate for the rate over such barrier by $\mathcal{R}'(40) = e^{(\Delta \Phi - \Delta \Phi')/k_bT} \mathcal{R}(40) = 2.2 \times 10^{-5} < \mathcal{R} = 5.1 \times 10^{-5}$ which is plotted as a purple dot in Fig. \ref{fig:rates}. This rate is closer to the rate by direct simulations $\mathcal{R}_\mathrm{MD}(40) = 1.42 \times 10^{-5}$.

\section{Discussion}

The results presented here show how increased confinement of the external potential, going from the one-dimensionally to a two-dimensionally confining potential, affects the escape rate of the polymers, as shown in Fig. \ref{fig:rates1d2d}. The added confinement lowers the number of possible configurations that a self-avoiding chain can take when it is sitting in the minimum, increasing the free energy of the initial state with respect to the transition state. The curvature of the external potential in the added dimension, the $y$-direction, is the same at the barrier and at the initial state minimum, and thus this effect is relatively more important in the initial state since the polymer tends to be elongated at the barrier. 

A clear minimum in the escape rate of self-avoiding polymers is obtained for intermediate length, about $N=30$, and this becomes even more pronounced in the two-dimensional case. The escape rate of ideal polymers, where repulsive interaction between the non-adjacent beads has been turned off, does not show such a minimum for the range studied here. This is consistent with the interpretation that the crowding of the beads in the self-avoiding polymers in the initial state well is responsible for lowering the free energy barrier for escape.  In the case of ideal polymers, such crowding effects are largely absent since only adjacent beads are subject to a repulsive interaction.

Qualitatively correct trends are obtained by applying Kramers rate theory to a one-dimensional reaction coordinate defined as the $x$-component of the centroid coordinate. However, the escape rate is overestimated for the longer polymers. For the ideal polymers, Kramers rate theory gives closer agreement with the PIHD simulations. This overestimate of the rate for the self-avoiding polymers can also be somewhat reduced by defining the effective potential curve as a thermal average of the external potential only, $V_{\mathrm{eff}}(x_\mathrm{C},N)  = \langle V_\mathrm{ext} \rangle_\mathrm{C}$. The results are shown in Fig. \ref{fig:rates}. This has almost no effect on the rate estimated for the ideal polymers, but significantly reduces that of the longer self-avoiding polymers, essentially through cancellation of errors.

The results presented here illustrate that a centroid coordinate cannot give a good reaction coordinate for the longer polymers, as also concluded by Debnath and coworkers \cite{Debnath2010}. A similar problem in defining a one dimensional reaction coordinate has been discussed in the context of the polymer reversal problem \cite{Huang2008, Zheng2011}. A good reaction coordinate needs to include information about the shape as well as the location of the polymer at the transition state. This result is similar to what has been concluded in quantum mechanical rate theory where Feynman path integrals are used to represent quantum delocalization.  Here, more beads need to be introduced in the path integrals the lower the temperature becomes, so an analogy exists between reduced temperature in the tunneling problems and length of polymers in the classical polymer escape problems.  Calculations of tunneling rates using the centroid as reaction coordinate have, indeed, shown an unphysical increase in tunneling rate as temperature is lowered, see for example Ref. \cite{Mills1997} and a good quantum transition state needs to be defined in terms of both location and shape of the Feynman paths \cite{Mills1998}. In the present case, a linear combination of the centroid coordinate and radius of gyration could be used as a good reaction coordinate, but the proper combination of the two will depend on the length of the polymer.  A systematic optimization of the location and orientation of a hyper-planar dividing surface so as to maximize the transition state  free energy could possibly be used for the polymer escape problem, analogous to what has been done for diffusion problems \cite{Johannesson2001,Bligaard2005}, It may also turn out that a hyper-planar dividing surface does not provide sufficient flexibility to confine the longer polymers to the bottleneck region for the escape.  This will be studied in future work. 

\section*{Acknowledments}

This work was supported by the Academy of Finland through the FiDiPro program (grant no. 263294) and the COMP CoE grant (no. 251748). The numerical calculations were carried out at the CSC -- IT Center for Science Ltd in Espoo, Finland.


\end{document}